\documentclass[aps,pra,showkeys,twocolumn,superscriptaddress,groupedaddress]{revtex4}
\usepackage{graphicx}
\usepackage{bm}% bold math
\usepackage{amsmath}% equation formating
\usepackage{bbold}
\usepackage[utf8]{inputenc}
\usepackage{color}
\usepackage{soul}
\usepackage[colorlinks=true, citecolor=blue]{hyperref}

\begin{document}
\title{Quantum simulation of ultrastrongly coupled bosonic modes using superconducting circuits}

\author{S. Fedortchenko}
\email{serguei.fedortchenko@univ-paris-diderot.fr}
\affiliation{Laboratoire Mat\' eriaux et Ph\' enom\`enes Quantiques, Sorbonne Paris Cit\' e, Universit\' e Paris Diderot, CNRS UMR 7162, 75013, Paris, France}

\author{S. Felicetti}
\affiliation{Laboratoire Mat\' eriaux et Ph\' enom\`enes Quantiques, Sorbonne Paris Cit\' e, Universit\' e Paris Diderot, CNRS UMR 7162, 75013, Paris, France}

\author{D. Markovi\'c}
\affiliation{Laboratoire Pierre Aigrain, \'Ecole Normale Sup\'erieure-PSL Research University, CNRS, Universit\'e Pierre et Marie Curie-Sorbonne Universit\'es, Universit\'e Paris Diderot-Sorbonne Paris Cit\'e, 24 rue Lhomond, 75231 Paris Cedex 05, France}

\author{S. Jezouin}
\affiliation{Laboratoire Pierre Aigrain, \'Ecole Normale Sup\'erieure-PSL Research University, CNRS, Universit\'e Pierre et Marie Curie-Sorbonne Universit\'es, Universit\'e Paris Diderot-Sorbonne Paris Cit\'e, 24 rue Lhomond, 75231 Paris Cedex 05, France}

\author{A. Keller}
\affiliation{Laboratoire Mat\' eriaux et Ph\' enom\`enes Quantiques, Sorbonne Paris Cit\' e, Universit\' e Paris Diderot, CNRS UMR 7162, 75013, Paris, France}
\affiliation{Institut des Sciences Mol\'eculaires d'Orsay (ISMO), CNRS, Univ. Paris Sud, Universit\'e Paris-Saclay, F-91405 Orsay (France)}

\author{T. Coudreau}
\affiliation{Laboratoire Mat\' eriaux et Ph\' enom\`enes Quantiques, Sorbonne Paris Cit\' e, Universit\' e Paris Diderot, CNRS UMR 7162, 75013, Paris, France}

\author{B. Huard}
\affiliation{Laboratoire Pierre Aigrain, \'Ecole Normale Sup\'erieure-PSL Research University, CNRS, Universit\'e Pierre et Marie Curie-Sorbonne Universit\'es, Universit\'e Paris Diderot-Sorbonne Paris Cit\'e, 24 rue Lhomond, 75231 Paris Cedex 05, France}
\affiliation{Laboratoire de Physique, \'Ecole Normale Supe\'erieure de Lyon, 46 all\'ee d'Italie, 69364 Lyon Cedex 7, France}

\author{P. Milman}
\affiliation{Laboratoire Mat\' eriaux et Ph\' enom\`enes Quantiques, Sorbonne Paris Cit\' e, Universit\' e Paris Diderot, CNRS UMR 7162, 75013, Paris, France}

\begin{abstract}
The ground state of a pair of ultrastrongly coupled bosonic modes is predicted to be a two-mode squeezed vacuum. However, the corresponding quantum correlations are currently unobservable in condensed matter where such a coupling can be reached, since it cannot be extracted from these systems. Here, we show that superconducting circuits can be used to perform an analog simulation of a system of two bosonic modes in regimes ranging from strong to ultrastrong coupling. More importantly, our quantum simulation setup enables us to detect output excitations that are related to the ground-state properties of the bosonic modes. We compute the emission spectra of this physical system and show that the produced state presents single- and two-mode squeezing simultaneously.
\end{abstract}

\vskip2pc

\vskip2pc 
\maketitle

\section{Introduction}
\label{Introduction}

Theoretically predicted more than a decade ago for two-dimensional electron gases~\cite{Ciuti}, and later for superconducting circuits \cite{Bourassa}, ultrastrong coupling is a fascinating regime of light-matter interaction. In strong coupling, quantum systems are coupled at a higher rate than any dissipation process, while in ultrastrong coupling they are coupled at a rate that is non-negligible even compared to the dynamics of each system taken separately. As a consequence, the rotating-wave approximation cannot be performed and all the terms of the coupling Hamiltonian should be, \textit{a priori}, considered. These terms play an important role in the properties of the system \cite{Ciuti,Ciuti2}, and the number of excitations is not conserved throughout the dynamics. Hence, the ground state is deeply modified. In recent years, this regime has been experimentally achieved in various physical systems: first, in cavity-embedded semiconductor quantum wells \cite{Gunter,Anappara,Todorov}, where the ultrastrong coupling was originally predicted, as well as in superconducting circuits \cite{Niemczyk,FornDiaz,FornDiaz2,Yoshihara1,Yoshihara2}, and in cavities confining molecules \cite{Schwartz,KenaCohen,George}.

Quantum simulation of ultrastrong coupling, or even deep strong coupling \cite{Casanova}, has recently received growing interest as the only way to probe dynamics or exotic features that are currently out of reach in genuine physical systems. In analog quantum simulators, the implementation relies on properly driven strongly interacting systems that effectively behave as ultrastrongly coupled modes and exhibit the corresponding characteristic features. Previous theoretical works have focused on simulating the interaction between a qubit and a cavity mode, namely, the quantum Rabi model, in several physical systems, with, for instance, proposals in light transport in photonic superlattices \cite{Longhi}, in superconducting circuits \cite{Ballester,Mezzacapo,Lamata}, in cavity quantum electrodynamics \cite{Grimsmo}, in trapped ions \cite{Pedernales}, and in ultracold atoms where the first and second Bloch bands in the first Brillouin zone encode the qubit \cite{Felicetti}. On the experimental side, quantum simulations of the Rabi model in the ultrastrong and deep strong coupling regimes have been reported in photonic superlattices \cite{Crespi}, and in superconducting circuits \cite{Langford,Braumuller}.

Surprisingly, the important case of a quantum simulation of ultrastrongly coupled bosonic modes is still missing. Remarkably, the ultrastrong interaction between two bosonic modes has the particularity of producing a two-mode squeezed vacuum in the ground state \cite{Ciuti,Ciuti2}. However, this squeezed state cannot lead to actual excitations coming out of the system and, thus, cannot be directly observed. In the case of spin-boson ultrastrong coupling, methods to probe the ground-state properties of the system were proposed in Refs.~\cite{Lolli,Peropadre,Cirio}. Nonetheless, in the case of two ultrastrongly coupled bosonic modes, the only studied solution to the problem is to modulate the coupling between the two modes in time~\cite{DeLiberato,Fedortchenko}. However, measuring the corresponding correlations between the two output channels of both bosonic modes seems currently out of reach for physical implementations of light-matter coupling~\cite{Gunter,Anappara,Todorov} since matter excitations (for instance the electron gas of quantum wells) decay through a nonradiative channel. Because of these serious experimental issues, a quantum simulation of ultrastrongly coupled bosonic modes is timely.

In this paper, we propose a way to realize it using the three-wave mixing process of a superconducting device. The Josephson mixer is made of a ring of four Josephson junctions that couples two microwave resonators~\cite{Bergeal,Abdo2}. It has been demonstrated to act as a microwave amplifier near the quantum limit \cite{Bergeal2,Roch}, to generate two-mode squeezed vaccum shared between two traveling or stationary microwave modes \cite{Flurin, FlurinQnode}, to realize coherent frequency conversion \cite{Abdo} and to act as a circulator or directional amplifier \cite{Sliwa}. The versatility of the Josephson mixer and the ease of measuring its two output channels make it an ideal platform to perform quantum simulations. Of particular relevance to our goal, it enables us to fully characterize the squeezing of its two output transmission lines. We propose here to drive this device in such a way that, in a particular rotating frame, its effective Hamiltonian is formally equivalent to the boson-boson ultrastrong coupling Hamiltonian. The peculiar properties of the simulated ground state lead to squeezing of the physically observable output modes in the laboratory frame. We predict the emission of an unusual two-mode output state, where both modes exhibit single-mode squeezing and, additionally, have quantum correlations between them.

The paper is organized as follows. First, we briefly characterize the squeezing properties of two ultrastrongly coupled bosonic modes in the ground state. In particular, we show that this ground state differs from a two-mode squeezed vacuum. Next, we make our model explicit, showing how driving the Josephson mixer leads to an effective Hamiltonian that simulates ultrastrong coupling. Then we predict the squeezing in the emission spectrum of the system, with realistic parameters. Finally, we discuss the physical meaning of the emitted squeezing, in particular, by interpreting the results in terms of two equivalent models, each one having a different environment.

\section{Ground state squeezing}
\label{GSsqueezing}

Let us start by studying the squeezing of a pair of ultrastrongly coupled bosonic modes in their ground state. For this we introduce the light-matter Hamiltonian we want to simulate,
\begin{equation}
\hat{H} = \omega_{\alpha} \hat{a}^{\dagger} \hat{a} + \omega_{\beta} \hat{b}^{\dagger} \hat{b}  +   G (\hat{a} + \hat{a}^{\dagger}) (\hat{b} + \hat{b}^{\dagger}),
\label{HUSC}
\end{equation}
where $\hat{a}$ ($\hat{b}$) and $\omega_\alpha$ ($\omega_\beta$) are the annihilation operator and the frequency of the light (matter) mode, and $\hbar=1$. The two modes are coupled at a rate $G$. We consider here the Hamiltonian (\ref{HUSC}) in its most elementary form. For instance, we do not include extra terms such as a squared electromagnetic vector potential, as is the case in semiconductors described in the Coulomb gauge \cite{Ciuti,Ciuti2}. Indeed, while the versatility of superconducting circuits would allow us to simulate extra terms, we choose to restrict the simulation to the simplest  form of ultrastrong coupling in the present paper \cite{footnote1}.

In order to identify the ground state of the Hamiltonian (\ref{HUSC}), we first apply the Hopfield method \cite{Hopfield} to identify the two eigenmodes of the system, which are called polaritons in case of a genuine light-matter interaction. The annihilation operators $\hat{p}_1$ and $\hat{p}_2$ of the two eigenmodes are expressed \cite{Ciuti,Ciuti2} as linear combinations of  $\hat{a}, \hat{b}, \hat{a}^\dagger$ and $\hat{b}^\dagger$; this is a Gaussian operation. Their expressions as well as their eigenvalues determine the validity of this model (see Appendix \ref{AppendixValidity}). We then express $\hat{a}$ and $\hat{b}$ as a function of the eigenmode operators $\hat{p}_1$ and $\hat{p}_2$. The ground state $\vert \text{GS} \rangle$ being defined as $\hat{p}_1 \vert \text{GS} \rangle =\hat{p}_2 \vert \text{GS} \rangle=0$, we can fully characterize the squeezing of the original modes $a$ and $b$ in the ground state $\vert \text{GS} \rangle$  by computing the covariance matrix $\mathcal{V}=\{\langle  x_ix_j + x_jx_i \rangle_{\vert \text{GS} \rangle}-2\langle x_i \rangle_{\vert \text{GS} \rangle}\langle x_j \rangle_{\vert \text{GS} \rangle}\}_{ij}$ in the basis $\{x_1,x_2,x_3,x_4\}=\{(\hat{a}^\dagger+\hat{a})/\sqrt{2},(i\hat{a}^\dagger-i\hat{a})/\sqrt{2},(\hat{b}^\dagger+\hat{b})/\sqrt{2},(i\hat{b}^\dagger-i\hat{b})/\sqrt{2}\}$ \cite{Serafini,Adesso}.

\begin{figure}[!t]
\centering
\includegraphics[width=0.48\textwidth]{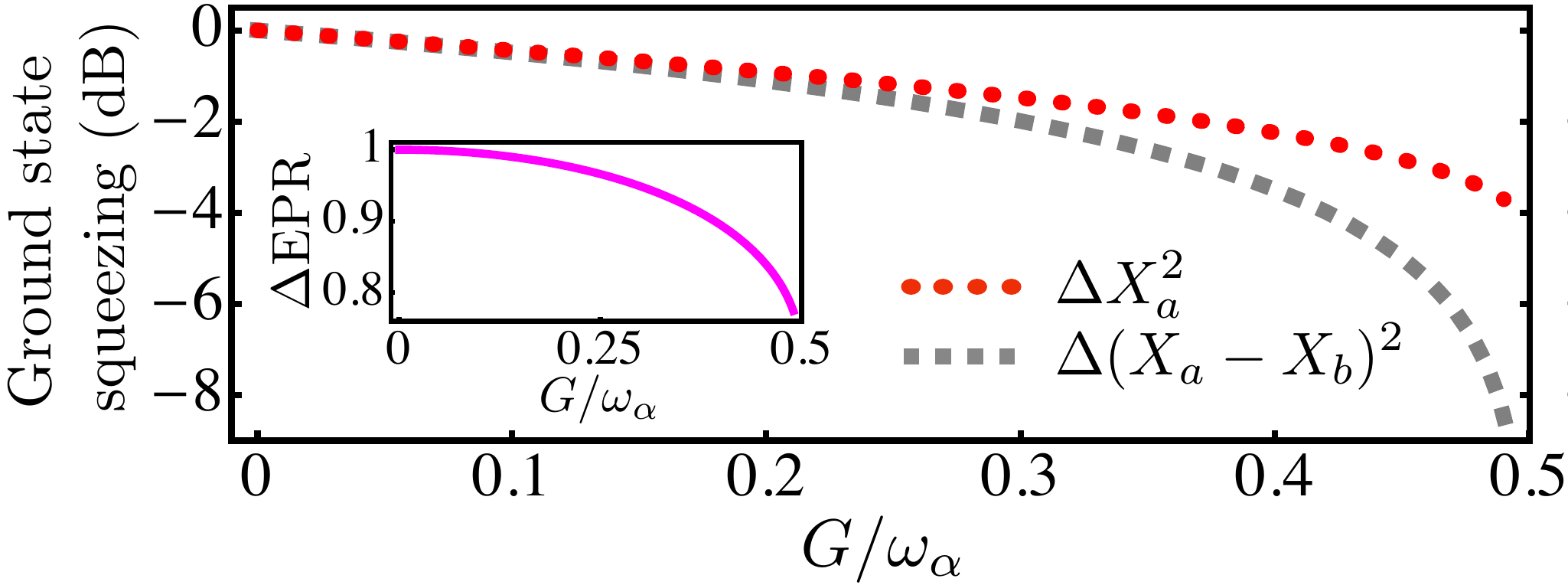}
\caption{Squeezing of ultrastrongly coupled $a$ and $b$ modes in the ground state, as a function of the coupling rate $G$. Red dotted line: single mode squeezing of the quadrature $\hat{X}_{a}$. Here the modes are degenerate $\omega_\alpha=\omega_\beta$, therefore the squeezing of $\hat{X}_b$ is the same as of $\hat{X}_{a}$. Gray dashed line: two-mode squeezing of the collective quadrature $\hat{X}_a-\hat{X}_b$. Inset: EPR variance $\Delta\text{EPR}= \Delta (X_a-X_b)^2 + \Delta (Y_a+Y_b)^2$.}
\label{FigureGroundState}
\end{figure}

In Fig.~\ref{FigureGroundState}, we show the single-mode squeezing, the two-mode squeezing, and the EPR variance (a measure of entanglement) in the ground state of a pair of ultrastrongly coupled bosonic modes, as a function of the coupling constant $G$. Specifically, we show the squeezing in dB using the following logarithmic scale $S_{X_{\theta}} = 10 \log_{10}{( \langle \Delta \hat{X}_{\theta}^2 \rangle / \langle \Delta \hat{X}_{\text{vac}}^2 \rangle)}$, where $\langle \Delta \hat{X}_{\text{vac}}^2 \rangle=1/2$ corresponds to the noise of a vacuum state. We use the definitions $\langle \Delta \hat{X}_{\theta}^2 \rangle = \langle \hat{X}_{\theta}^2 \rangle - \langle \hat{X}_{\theta} \rangle^2$,  $\hat{X}_{\theta} = (e^{-i \theta} \hat{a} + e^{i \theta} \hat{a}^\dagger)/\sqrt{2}$, with $\hat{X}_{\theta=0}=\hat{X}$ and $\hat{X}_{\theta=\pi/2}=\hat{Y}$. One can note that the ground state shows a significant amount of squeezing in the single-mode picture, as well as in the two-mode picture, enough to be detected by a Gaussian entanglement witness: the EPR variance goes below 1 \cite{Duan,Simon}. Note that since here the two modes are at resonance, only $\Delta \hat{X}_a^2$ is shown, because $\Delta \hat{X}_b^2$ has exactly the same amount of squeezing. We thus verified that there is two-mode squeezing in the ground state, and additionally found single-mode squeezing as well. Note that it is possible to intuitively predict squeezing in the Hamiltonian~\eqref{HUSC}, by rewriting it in terms of particular collective operators (see Appendix \ref{AppendixSqueezingH}).

\begin{figure*}[!t]
\centering
\includegraphics[width=0.7\textwidth]{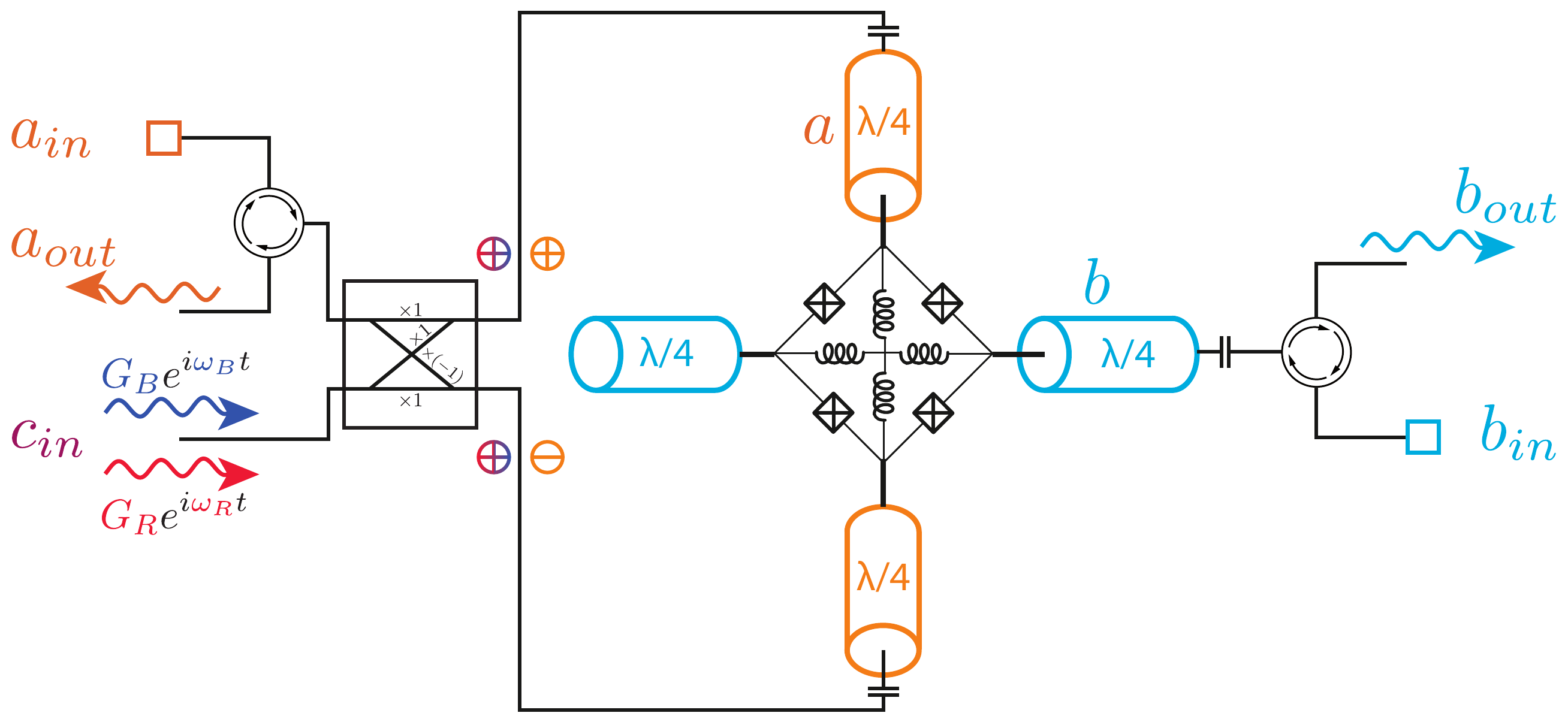}
\caption{Scheme of a possible implementation based on a Josephson mixer~\cite{Abdo2}. A ring of four Josephson junctions is shorted by inductors and couples two $\lambda/2$ microwave resonators of frequency $\omega_a$ and $\omega_b$. Capacitors couple the resonators to transmission lines leading to decay rates $\gamma_a+\gamma_L$ and $\gamma_b+\gamma_L$, where $\gamma_L$ corresponds to internal losses of the resonators. This circuit implements three-wave mixing between the nondegenerate modes $a$ and $b$ and a mode $c$ that can be addressed using a signal driven with the same phase on each port of the resonator $a$. One may use a $180^\circ$  hybrid coupler (box on the left) to selectively couple $a$ and $c$ modes to two separate transmission lines. Circulators ensure that the input modes $a_\mathrm{in}$ and $b_\mathrm{in}$ are prepared in the vacuum state by thermalizing a $50~\Omega$ load at $T\ll \hbar\omega_{a,b}/k_B$. When mode $c$ is driven off resonance by two tones at frequency $\omega_B=\omega_a+\omega_b+2\delta$ and $\omega_R=\omega_a-\omega_b$, it reproduces the physics of two ultrastrongly coupled bosonic modes of frequency $\delta$. Signatures of the ultrastrong coupling can be observed in the squeezing properties of the noise in ports $a_\mathrm{out}$ and $b_\mathrm{out}$. }
\label{FigureJPC}
\end{figure*}

\section{Modeling the quantum simulation}
\label{Model}

In this section we present the model of our quantum simulation, which could be used as a tool to measure the squeezing of the output field extracted from the system in its ground state. As mentioned in previous sections, our model is based on the Josephson mixer (Fig.~\ref{FigureJPC}) \cite{Abdo2}, where the interaction Hamiltonian of the three-wave mixing process reads,
\begin{equation}
\hat{H}_{\text{int}} =   \chi ( \hat{c} + \hat{c}^{\dagger}) (  \hat{a} + \hat{a}^{\dagger} ) ( \hat{b} + \hat{b}^{\dagger}),
\label{H3WM}
\end{equation}
where $\hat{c}$, $\hat{a}$, and $\hat{b}$ are the annihilation operators of the spatially separated pump, $a$ and $b$ microwave modes, respectively. This purely three-wave mixing Hamiltonian is close to what the circuit in Fig.~\ref{FigureJPC} can realize for a well chosen value of the magnetic flux threading the inner loops of the Josephson ring. For more details on how to obtain the system interaction Hamiltonian \eqref{H3WM} from the general Hamiltonian describing the Josephson mixer, as well as on the measurement process of the outputs of modes $a$ and $b$, we refer the reader to Ref.~\cite{FlurinThesis}.

To generate an effective Hamiltonian that is formally equivalent to Eq. (\ref{HUSC}), we drive the system with a two-tone radiation. A blue pump drives mode $c$ with an amplitude $c_B$ at frequency $\omega_{B} = \omega_{a} + \omega_{b} + 2 \delta$, while a red pump drives the same mode $c$ with an amplitude $c_R$ at $\omega_{R} = \omega_{a} - \omega_{b}$. Here $\omega_{a}$ and $\omega_{b}$ are the frequencies of modes $a$ and $b$, and  $2 \delta$ is a small detuning compared to them. Mode $c$ being driven off resonance, we use the stiff pump approximation and describe its amplitude as a complex number instead of an operator. The interaction Hamiltonian now has two three-wave mixing terms, which result in the following effective Hamiltonian in the rotating frame where mode $a$ rotates at $\omega_a + \delta$ and mode $b$ at $\omega_b + \delta$ (see Appendix \ref{AppendixEffectiveH}),
\begin{equation}
\hat{H}_{\text{eff}}  =  \delta \, \hat{a}^{\dagger} \hat{a} + \delta \, \hat{b}^{\dagger} \hat{b} + G_{B} (  \hat{a}^\dagger  \hat{b}^\dagger  + \hat{a} \,  \hat{b})  + G_{R} ( \hat{a}^{\dagger} \hat{b} + \hat{a} \,  \hat{b}^\dagger ),
\label{Heff}
\end{equation}
where $G_{B,R}=\chi c_{B,R}$ is time-independent and results from the physical time-dependent coupling rate $\tilde{G}_{B,R}(t) = G_{B,R} e^{-i \omega_{B,R} t}$. The derivation above is valid only for low three-wave mixing rates $\vert G_{B,R} \vert \ll \omega_a, \omega_b, \vert \omega_a - \omega_b \vert$. In the case when $G_R=0$, the Hamiltonian describes parametric amplification, which results in two-mode squeezing, while when $G_B=0$, it describes a beam splitter between modes $a$ and $b$. Now if $G_B = G_R=G$, Eq.~(\ref{Heff}) has exactly the same form as Eq.~(\ref{HUSC}) if $\omega_\alpha=\omega_\beta$. Here, $\delta$ plays the role of the bosonic mode free oscillation frequency. It is now clear that when the coupling $G$ becomes comparable to $\delta$, the doubly pumped Josephson mixer simulates ultrastrongly coupled modes, even if the genuine coupling is much smaller than the genuine free oscillation frequencies of the physical system. It is worthwhile to note that although the simulated coupling rates $G_{B,R}$ are time-independent, as in the case of genuine ultrastrong coupling in semiconductors \cite{Gunter,Todorov,Anappara}, the actual coupling rate oscillates in the laboratory frame of the output ports of modes $a$ and $b$. Note that a method to obtain a genuine ultrastrong coupling between two bosonic modes in superconducting circuits was proposed in Ref.~\cite{Peropadre2}. There the coupling is mediated not by a third oscillator but by a SQUID, and while the physical coupling could in principle reach the ultrastrong regime, its predicted coupling-to-frequency ratio does not reach the highest values of the coupling–to–effective frequency ratio leading to the interesting squeezing properties that we study here and that are realistically achievable in our quantum simulation.

\section{Results: emission spectra of the system}
\label{Results}

\begin{figure}[!t]
\centering
\includegraphics[width=0.47\textwidth]{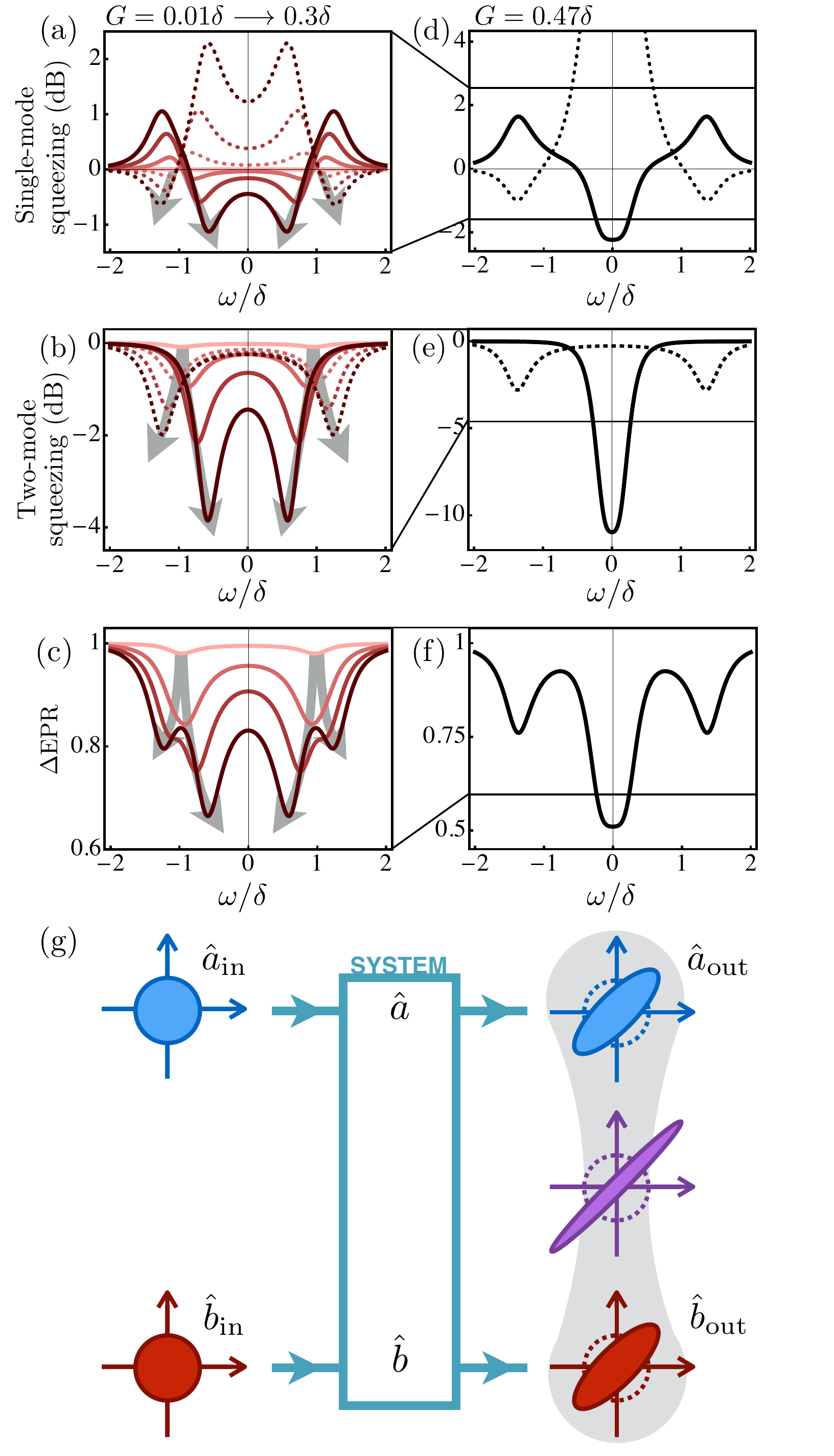}
\caption{Noise spectra of the system output for $G_B=G_R=G$. (a), (d) Noise spectra of $\hat{X}_a$ (dotted line) and of $\hat{Y}_a$ (full line). (b), (e) Noise spectra of $\hat{X}_a - \hat{X}_b$ (dashed line) and of of $\hat{Y}_a + \hat{Y}_b$ (full line). (c), (f) EPR variance $\Delta\text{EPR}= \Delta (X_a-X_b)^2 + \Delta (Y_a+Y_b)^2$. The parameters are: $\omega_a= 2 \pi \times 9$ GHz ; $\omega_b= 2 \pi \times 6$ GHz ; $\delta= 2 \pi \times 50$ MHz ; $\gamma_a=\gamma_b= 2 \pi \times 25$ MHz ; $\gamma_L= 2 \pi \times 0.5$ MHz. (a-c) Color code: each color or shade is associated with a value of $G/\delta$: lighter curves represent $G/\delta=0.01$; darker ones, $G/\delta=0.3$. From lightest to darkest curve, $G/\delta$ takes the values $\{  0.01, 0.1, 0.2, 0.3 \}$. Arrows follow the splitting of the resonance frequency as $G/\delta$ increases, simulating Rabi splitting. (d-f) Here we fix $G/\delta=0.47$, for which the two central dips (peaks) predicted in (a-c) merge. The black horizontal lines indicate  the plot range in (a-c). (g) Summary of the generation of the squeezed output state represented as contours of the marginals of the Wigner function in single- and two-mode quadrature phase spaces. }
\label{FigureOutputSpectra}
\end{figure} 

%(a)-(f) $G=G_B=G_R$, and from lighter to darker shades, $G/\delta$ takes the values $\{  0.01, 0.1, 0.2, 0.3, 0.47 \}$. (d)-(f) The black horizontal lines indicate  the plot range in (a)-(c)

We now show the expected results of the quantum simulation, by determining the radiation emitted by the device in the regime where the latter can be described by the effective Hamiltonian~(\ref{Heff}). Each mode $a$ or $b$ is connected to a transmission line at a rate $\gamma_{a,b}$ and is subject to internal losses at a rate $\gamma_L$. In the input-output formalism \cite{Ciuti2,Gardiner,GardinerBook,WallsBook}, we are interested in the state of the output modes whose operators are $\hat{a}_{\text{out}}$ and $\hat{b}_{\text{out}}$. They are related to the input mode operators by the input-output relations  $\hat{a}_{\text{out}} = \hat{a}_{\text{in}} + \sqrt{\gamma_a} \hat{a}$ and $\hat{b}_{\text{out}} = \hat{b}_{\text{in}} + \sqrt{\gamma_b} \hat{b}$. From the known input state, one gets the output state from the above expressions and from the quantum Langevin equations for the intracavity operators
\begin{align}
 \dot{\hat{a}} (t)  = & - i \delta \hat{a} (t) - \frac{\gamma_a + \gamma_L}{2} \hat{a} (t) - i G \big( \hat{b} (t) + \hat{b}^\dagger (t) \big) \nonumber \\
&  -\sqrt{\gamma_a} \hat{a}_{\text{in}} (t)  -\sqrt{\gamma_L} \hat{f}_{a} (t) \label{equationOfmotion1} \\
 \dot{\hat{b}} (t)  = & - i \delta \hat{b} (t) - \frac{\gamma_b + \gamma_L}{2} \hat{b} (t) - i G \big( \hat{a} (t) + \hat{a}^\dagger (t) \big) \nonumber \\
&  -\sqrt{\gamma_b} \hat{b}_{\text{in}} (t)  -\sqrt{\gamma_L} \hat{f}_{b} (t).
\label{equationOfmotion2}
\end{align}
where $\hat{f}_a$ and $\hat{f}_b$ are noise operators modeling the internal losses of the system.
It is straightforward to solve these equations in the frequency domain to obtain the expressions of $\hat{a}_{\text{out}} [\omega]$ and $\hat{b}_{\text{out}} [\omega]$, from which we obtain the covariance matrix $\mathcal{V}$ that fully characterizes the Gaussian output state. The noise properties are directly given by the elements of $\mathcal{V}$.

In Figs.~\ref{FigureOutputSpectra}(a)-(f) we show the output noise spectra of single-mode quadratures $\hat{X}_a$ and $\hat{Y}_a$, of two-mode quadratures $\hat{X}_a-\hat{X}_b$ and $\hat{Y}_a+\hat{Y}_b$, and the EPR variance \cite{Duan,Simon}, as a function of frequency. In the rotating frame, a signal at frequency $\omega$ corresponds to $\omega_a + \delta + \omega$ for mode $a$, and to $\omega_b + \delta + \omega$ for mode $b$ in the laboratory frame. We do not show the noise spectra of $\hat{X}_b$ and $\hat{Y}_b$ since they are the same as for $\hat{X}_a$ and $\hat{Y}_a$, both modes having the same effective frequency $\delta$, and the dissipation rates $\gamma_{a}$ and $\gamma_b$ being assumed identical. As expected the output radiation is more squeezed for stronger coupling $G=G_B=G_R$. Furthermore, the squeezing becomes visible in the figures when ultrastrong coupling is reached for $G \gtrsim 0.1 \delta$. The behavior of the system in the physical implementation picture is illustrated in Fig.~\ref{FigureOutputSpectra}(g). When modes $a$ and $b$ are in the vacuum state at the input, the output of the system is in an unusual two-mode state, where each mode is squeezed, while the two modes are quantum correlated, similarly to the ground state of the Hamiltonian (\ref{HUSC}), shown in Fig.~\ref{FigureGroundState}. Interestingly, the squeezing we predict here occurs between two propagating modes that are separated both in space and frequency.

Let us now comment on the shape of the spectra. In the rotating frame, we show the positive and negative parts of the frequency spectrum, which correspond to measurable noise powers at positive frequencies in the laboratory frame. In Figs.~\ref{FigureOutputSpectra}(a)-(c), we can see that for the smallest shown coupling $G=0.01\delta$, the spectra develop a resonance at $\omega=\pm\delta$, symmetrically for positive and negative frequencies. This resonance occurs at the transition frequency $\delta$ of the effectively degenerate modes $a$ and $b$ in the rotating frame (see Eq. (\ref{Heff})). As $G$  increases, the resonance splits into two, leading to four dips in the EPR variance Figs.~\ref{FigureOutputSpectra}(c). This can be understood as the vacuum Rabi splitting of both effective modes, as already observed in a physically ultrastrongly coupled light-matter system~\cite{Gunter}.

As the splitting increases with $G$, one of the two resonance frequencies resulting from the Rabi splitting shifts towards $\omega=0$. In Figs~\ref{FigureOutputSpectra}(a)-(c), this can be seen as two dips getting closer to the origin, corresponding to the resonance frequency and its image on the negative part of the spectrum. When $G \approx 0.5 \delta$, the dips merge at the origin and the resonance occurs at $\omega=0$. This is shown in Figs.~\ref{FigureOutputSpectra}(d)-(f), where there are no longer four dips but only three, and the one at the origin shows the largest amount of two-mode squeezing. Thereby, the EPR variance almost reaches the lower bound of $0.5$, which corresponds to an optimal case where $\hat{Y}_a+\hat{Y}_b$ is infinitely squeezed, while $\hat{X}_a-\hat{X}_b$ is shot noise limited only.  Besides, the single mode squeezing in the quadratures $\hat{Y_a}$ and $\hat{Y_b}$ reaches almost $-3$~dB. This is in fact the maximal expected single-mode squeezing  one can hope for. We note that if the two outputs were combined in a 50:50 beam splitter (with frequency conversion on one arm), one of the output modes would be in an infinitely squeezed state while the other would be in the vacuum state \cite{Laurat}; the reverse has been demonstrated in \cite{Ku}. This can be done using an extra Josephson mixer as in Ref.~\cite{Flurin} but in frequency conversion mode. 

The squeezing amplitudes in Fig.~\ref{FigureOutputSpectra}(d)-(e) are limited by the realistic internal losses and coupling rates to the transmission lines we use in the model. Their minimal  value is set  by the need to stay in the regime where the three wave mixing Hamiltonian~(\ref{H3WM}) is valid~\cite{FlurinThesis}.  The figures stop at $G\approx 0.5\delta$ since beyond that point, the Hamiltonian \eqref{Heff} has no stable solution and extra terms should be included to make the Hamiltonian physically sound again (see Appendix \ref{AppendixValidity}). For instance, in case of the Dicke model modeling a spin ensemble coupled to a bosonic mode,  this value for the coupling is a critical point of a quantum phase transition~\cite{Emary,Emary2,Nataf}. In the proposed simulation using a Josephson mixer, these extra terms arise from a Taylor expansion of the Josephson Hamiltonian beyond second order.

It is worthwhile to wonder how realistic are the parameters we chose in Fig.~\ref{FigureOutputSpectra}. The phenomena we propose to observe require that $\gamma_L\ll \gamma_{a,b}<\delta$ and that $2G_{B,R}\lesssim\delta$. It is shown in Ref.~\cite{FlurinThesis} that 
\begin{equation}
\frac{2G_{B,R}}{\sqrt{\gamma_a\gamma_b}}\leq\frac{1}{4} \sqrt{\xi_a\xi_bQ_aQ_b},
\end{equation}
where $\xi<1$ is the participation ratio of the Josephson junction in the resonator \cite{FlurinThesis} and $Q$ is the quality factor of the resonator.
Therefore, in order to reach $2G_{B,R}\approx\delta$, one needs \begin{equation}
1<\frac{\delta}{\gamma_{a,b}}\leq\frac{1}{4} \sqrt{\xi_a\xi_bQ_aQ_b}.
\end{equation}
The condition that $\sqrt{\xi_a\xi_bQ_aQ_b}>4$ sets constraints on the device similar to the ones needed to realize a quantum limited amplifier using the Josephson mixer~\cite{Abdo2,Pillet} and is perfectly realistic. The parameters we chose in Fig.~\ref{FigureOutputSpectra} are thus within reach in standard devices.

\section{Discussion}
\label{Discussion}

\begin{figure}[!t]
\centering
\includegraphics[width=0.43\textwidth]{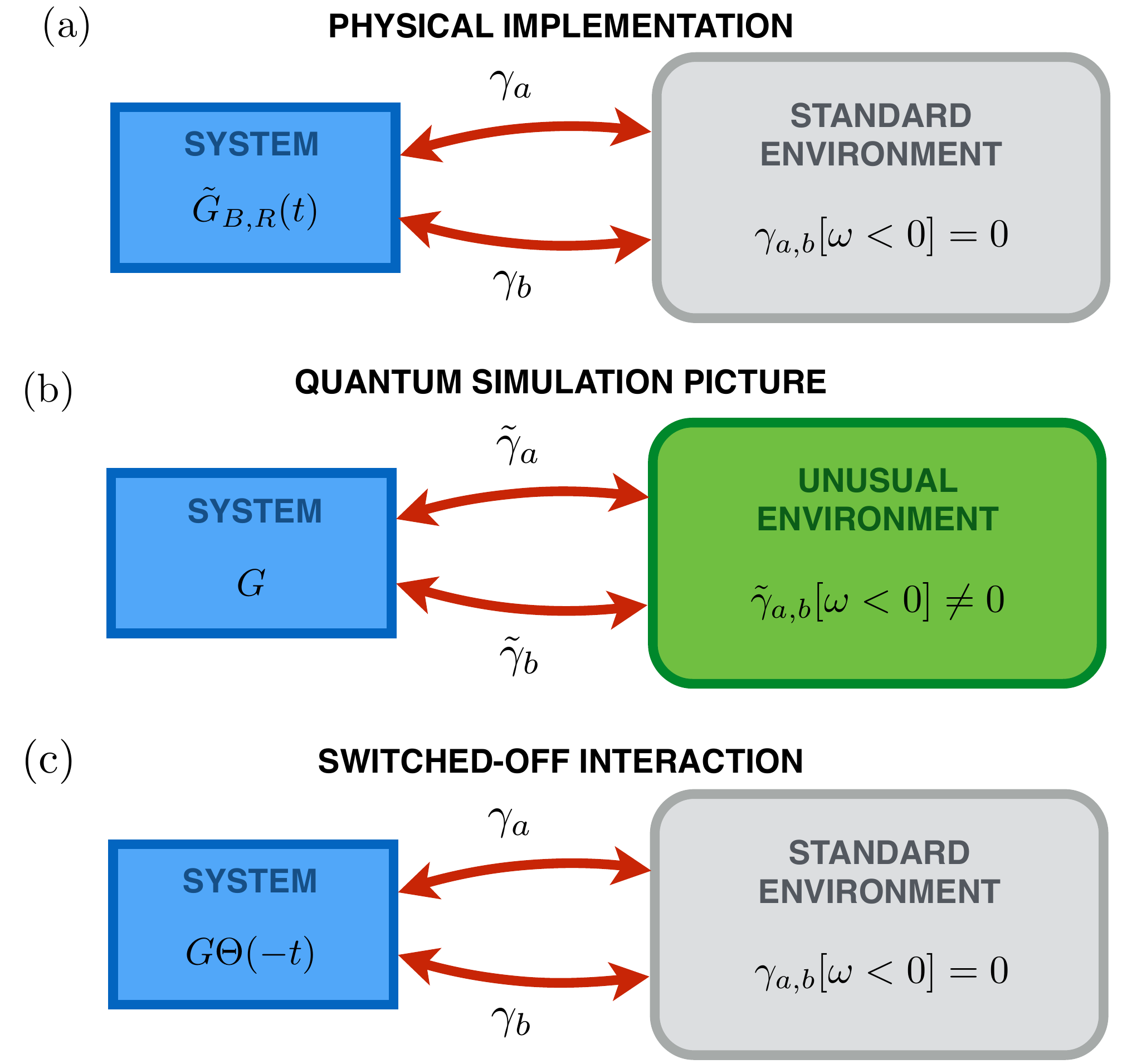}
\caption{Schemes of three models describing two interacting bosonic modes that are coupled to an environment, in which a squeezed field is emitted. (a) Physical picture with a time-dependent interaction and a standard environment, which is coupled at a rate that vanishes for negative frequencies. (b) Quantum simulated model mapped from (a), where the ultrastrong interaction is time independent and the baths are unusual with a support for $\gamma_{a,b}$ spanning positive and negative frequencies. (c) Genuine ultrastrongly interacting bosonic modes whose interaction is abruptly switched-off at time $0$ \cite{Ciuti}, and which are coupled to a standard environment. $\Theta (t)$ is the unit step function, introduced to model the switch-off of the interaction.}
\label{FigureDiscussion}
\end{figure}

So far, we have focused on the observable squeezing contained in the output modes of the physical system that simulates the ultrastrong coupling Hamiltonian~\eqref{Heff}. In this section, we interpret the nature of the output modes in the simulated picture. We summarize the key components of the physical system in Fig.~\ref{FigureDiscussion}(a). The coupling between modes $a$ and $b$ is modulated in time and each mode is coupled to a zero temperature bath at a rate $\gamma_{a,b}[\omega]$ with vanishing contribution from negative frequencies~\cite{Ciuti2}. In contrast, in the simulated picture, the modes are coupled at a fixed rate $G$ but interact with an unusual environment, whose coupling rates $\tilde{\gamma}_{a,b}[\omega]$ are nonzero at negative frequencies (Fig.~\ref{FigureDiscussion}(b)). This results from a shift of the zero frequency in the rotating frame of the simulation. The vacuum squeezing of the ultrastrongly coupled modes in their ground state can be understood as resulting from the excitations corresponding to the nonzero $\tilde{\gamma}_{a,b}[\omega]$ for $\omega<0$.

In order to release solely the ground-state photons, in genuine ultrastrongly coupled systems, one should abruptly switch off the interaction \cite{Ciuti,DeLiberatoThesis} [Fig.~\ref{FigureDiscussion}(c)]. In this way, the energy contained in the virtual excitations of the ground state is released from the cavity until the system reaches its new ground state, that of a noninteracting system. In our simulation, we can avoid turning off the interaction and still observe squeezing in the output, owing to the peculiarity of the environment [Fig.~\ref{FigureDiscussion}(b)].

\section{Conclusion}
\label{Conclusion}

In conclusion, we propose a superconducting circuit experiment to simulate ultrastrongly coupled bosonic modes. This work contributes to clarify their elusive ground-state properties. Using the high level of control of superconducting circuits enables us to access both modes directly, a feat that is not possible with current light-matter systems. Our proposal will determine the smooth transition from strong to ultrastrong coupling regimes by measuring the squeezing properties of the output modes. Beyond the fundamental interest in observing this transition, the unusual squeezing properties of the proposed device can be used as a resource for bath engineering~\cite{Aron,Aron2} and nonclassical state generation~\cite{Felicetti2,Stassi,Rossatto} in complex resonator networks.

\acknowledgements

The authors acknowledge Z. Leghtas, F. Portier, C. Ciuti, and P. Bertet for helpful discussions. This work was supported by the French Agence Nationale de la Recherche (ANR COMB project No. ANR-13-BS04-0014, GEARED project No. ANR-14-CE26-0018), by the Emergences program of Ville de Paris under grant Qumotel, by University Sorbonne Paris Cit\'e EQDOL, and by Paris Science et Lettres Idex PSL.

\appendix
\section{Validity of the model}
\label{AppendixValidity}

In this Appendix we briefly discuss the validity region of our model used to compute the ground state squeezing shown in Fig.~\ref{FigureGroundState}. To understand it we need the expressions of $\hat{p}_{1,2}$, the eigenmodes of the Hamiltonian~\eqref{HUSC}. These operators are linear combinations of $\hat{a}$ and $\hat{b}$,
\begin{equation}
\hat{p}_{1,2} = t_{1,2} \hat{a} + u_{1,2} \hat{b} + v_{1,2} \hat{a}^\dagger + w_{1,2} \hat{b}^\dagger,
\label{polaritons}
\end{equation}
where the coefficients $\vec{p}_{1,2}=\{ t_{1,2}, u_{1,2}, v_{1,2}, w_{1,2}  \}$ are obtained by diagonalizing the Hopfield matrix 
~\cite{Hopfield} for the Hamiltonian~\eqref{HUSC}. These coefficients are
\begin{align}
\vec{p}_{1} & =\frac{1}{\sqrt{N_1}}\left(
      \begin{array}{ccc}
         \frac{\sqrt{(\delta - 2 G)\delta}+ \delta}{G}-1  \\
         - \frac{\sqrt{(\delta - 2 G)\delta}+ \delta}{G}+1   \\
        -1 \\
        1
      \end{array} \right)
\label{p1} \\
\vec{p}_{2} & =\frac{1}{\sqrt{N_2}}\left(
      \begin{array}{ccc}
         \frac{\sqrt{(\delta + 2 G)\delta}+ \delta}{G}+1  \\
         \frac{\sqrt{(\delta + 2 G)\delta}+ \delta}{G}+1   \\
        1 \\
        1
      \end{array} \right)
\label{p2}
\end{align}
with eigenvalues
\begin{equation}
\omega_{1,2}=\sqrt{(\delta \mp 2G)\delta} \label{eigenvalues}.
\end{equation}
$N_{1,2}$ are the normalization coefficients, such that  the condition $\vert t_{1,2} \vert^2 + \vert u_{1,2} \vert^2 - \vert v_{1,2} \vert^2 - \vert w_{1,2} \vert^2 = 1$ is satisfied, imposed by the Bose commutation rule. With Eqs.~\eqref{p1} and \eqref{eigenvalues} one can see that when $G>\delta/2$, the model is not valid anymore.

\section{Two-mode squeezing operations}
\label{AppendixSqueezingH}

Let us show an intuitive picture in which the ground state squeezing is naturally predicted in a system described by the Hamiltonian~\eqref{HUSC}. We rewrite this Hamiltonian in terms of the following two collective modes $\hat{m}=(\hat{a} + \hat{b})/\sqrt{2}$ and $\hat{n}=(\hat{a} - \hat{b})/\sqrt{2}$, that are well defined bosonic modes,
\begin{align}
\hat{H} &= (\omega+G) \hat{m}^{\dagger} \hat{m} + (\omega-G)\hat{n}^{\dagger} \hat{n}  \nonumber \\
&+ \frac{G}{2} (\hat{m}^2 + (\hat{m}^{\dagger})^2) -\frac{G}{2} (\hat{n}^2 + (\hat{n}^{\dagger})^2),
\label{HUSCsqueez}
\end{align}
where we considered the case $\omega_\alpha=\omega_\beta=\omega$, a condition used in Fig.~\ref{FigureGroundState}. One can clearly see from Eq.~\eqref{HUSCsqueez} that modes $\hat{m}$ and $\hat{n}$ are independent and both ruled by a squeezing Hamiltonian. Hence their ground state is expected to be largely squeezed in the ultrastrong coupling regime. Note that the single-mode squeezing of $\hat{m}$ (resp. $\hat{n}$) implies a two-mode squeezing in the original $\hat{a},\hat{b}$ basis along $\hat{X}_a-\hat{X}_b$ (resp. $\hat{Y}_a+\hat{Y}_b$) as shown in Fig.~\ref{FigureGroundState}. However, while this alternative description in terms of modes $m$ and $n$ clearly shows the correlations between modes $a$ and $b$, it is less obvious to predict their single-mode squeezing, which can be definitely verified with the covariance matrix.

\section{Derivation of the effective Hamiltonian}
\label{AppendixEffectiveH}

Here we show the derivation of the effective Hamiltonian \eqref{Heff}. As mentioned in the main text, the interaction in the physical system is a three-wave mixing process between a pump mode $c$ and two microwave modes $a$ and $b$, described by Eq.~\eqref{H3WM}. However, since we drive the pump mode by a two-tone radiation, the interaction Hamiltonian now includes two three-wave mixing terms, and the full system Hamiltonian reads
\begin{align}
\hat{H} & =   \omega_{a} \, \hat{a}^{\dagger} \hat{a} + \omega_{b} \, \hat{b}^{\dagger} \hat{b} + \omega_{B} \, \hat{c}_{B}^{\dagger} \hat{c}_{B} + \omega_{R} \, \hat{c}_{R}^{\dagger} \hat{c}_{R}  \nonumber \\
& + \chi ( \hat{c}_B + \hat{c}^{\dagger}_B) (  \hat{a} + \hat{a}^{\dagger} ) ( \hat{b} + \hat{b}^{\dagger}) \nonumber \\
& + \chi ( \hat{c}_R + \hat{c}^{\dagger}_R) (  \hat{a} + \hat{a}^{\dagger} ) ( \hat{b} + \hat{b}^{\dagger}),
\label{H3WMtwotone}
\end{align}
In the interaction picture, this Hamiltonian reads
\begin{align}
\hat{H}_{\text{IP}} & = \chi (\hat{c}_{B} e^{-i \omega_{B} t} +  \hat{c}_{R} e^{-i \omega_{R} t}) ( \hat{a} \, \hat{b} \, e^{-i (\omega_{a} + \omega_{b}) t}  \nonumber \\
& + \hat{a} \,  \hat{b}^\dagger e^{-i (\omega_{a} - \omega_{b}) t}  +  \hat{a}^{\dagger} \hat{b} \,  e^{i (\omega_{a} -\omega_{b} ) t} \nonumber \\
& + \hat{a}^\dagger  \hat{b}^\dagger e^{i (\omega_{a} + \omega_{b}) t} )  + \text{h.c.},
\label{HTTip}
\end{align}
where the frequencies of the two-tone driving are 
\begin{align}
\omega_{B} & = \omega_{a} + \omega_{b}  + 2 \delta, \label{bluepump} \\
\omega_{R} & = \omega_{a} - \omega_{b}, \label{repump}
\end{align}
where $\vert \delta \vert \ll \omega_a, \omega_b, \vert \omega_a - \omega_b \vert$. Mode $c$ being driven off resonance, we use the stiff pump approximation and describe its amplitude as a complex number instead of an operator. Let us call $G_{B,R}=\chi c_{B,R}$ as the time independent parts of the coupling rates $\tilde{G}_{B,R}(t) = G_{B,R} e^{-i \omega_{B,R} t}$. Using Eqs.~\eqref{HTTip},\eqref{bluepump} and \eqref{repump}, we obtain
\begin{align}
\hat{H}_{\text{IP}} & = G_{B} ( \hat{a} \, \hat{b} \, e^{-2 i (\omega_{a} + \omega_{b} + \delta) t}  + \hat{a} \,  \hat{b}^\dagger e^{-2 i (\omega_{a} + \delta) t} \nonumber \\
& +  \hat{a}^{\dagger} \hat{b} \,  e^{-2 i (\omega_{b} + \delta) t} + \hat{a}^\dagger  \hat{b}^\dagger e^{-2 i \delta t} )  + \nonumber \\
& + G_{R} ( \hat{a} \, \hat{b} \, e^{-2 i \omega_{a} t}  + \hat{a} \,  \hat{b}^\dagger e^{-2 i (\omega_{a} - \omega_{b}) t} \nonumber \\
& +  \hat{a}^{\dagger} \hat{b} + \hat{a}^\dagger  \hat{b}^\dagger e^{2 i \omega_{b} t} ) + \text{h.c.},
\label{HTTip2}
\end{align}
We work in a regime where $\vert G_{B,R} \vert \ll \omega_a, \omega_b, \vert \omega_a - \omega_b \vert$ and $\vert G_{B,R} \vert \lesssim  \vert \delta \vert$, which allows us to perform a rotating wave approximation. Thus, in the interaction picture, the important terms that contribute to the evolution of the system are resonant in this rotating frame, or oscillate at $2\delta$, and all the other terms can be fairly neglected,
\begin{equation}
\hat{H}_{\text{IP}}  \approx G_{B} (  \hat{a}^\dagger  \hat{b}^\dagger e^{-2 i \delta t}  + \hat{a} \,  \hat{b} \, e^{2 i \delta t})  + G_{R} ( \hat{a}^{\dagger} \hat{b} + \hat{a} \,  \hat{b}^\dagger ).
\label{HTTip3}
\end{equation}
With a rather simple, yet judiciously chosen unitary transformation we obtain the effective Hamiltonian
\begin{equation}
\hat{H}_{\text{eff}}  =  \delta \, \hat{a}^{\dagger} \hat{a} + \delta \, \hat{b}^{\dagger} \hat{b} + G_{B} (  \hat{a}^\dagger  \hat{b}^\dagger  + \hat{a} \,  \hat{b})  + G_{R} ( \hat{a}^{\dagger} \hat{b} + \hat{a} \,  \hat{b}^\dagger ).
\label{HeffApdx}
\end{equation}
We are now in a rotating frame where mode $a$ oscillates at $\omega_{a} + \delta$ and mode $b$ oscillates at $\omega_{b} + \delta$. Any single mode squeezing or correlations observed in this frame at a frequency $\omega$ would correspond in the laboratory frame to $\omega_{a} + \delta + \omega$  and $\omega_{b} + \delta + \omega$ for modes $\hat{a}$ and $\hat{b}$ respectively.

\end{document}